\documentclass[a4paper,11pt]{article}
\usepackage{pos}

\title{Searching for time-dependent high-energy neutrino emission from X-ray binaries with IceCube}
 \ShortTitle{Neutrino Emission from X-ray Binaries}

\author{The IceCube Collaboration \\{\normalsize \normalfont(a complete list of authors can be found at the end of the proceedings)}}

\emailAdd{qliu@icecube.wisc.edu}

\abstract{X-ray binaries are long-standing source candidates of Galactic cosmic rays and neutrinos. The compact object in a binary system can be the site for cosmic-ray acceleration, while high-energy neutrinos can be produced by the interactions of cosmic rays in the jet of the compact object, the stellar wind, or the atmosphere of the companion star. We report a time-dependent study of high-energy neutrinos from X-ray binaries with IceCube using 7.5 years of muon neutrino data and X-ray observations. In the absence of significant correlation, we report upper limits on the neutrino fluxes from these sources and provide a comparison with theoretical predictions.

\vspace{4mm}
{\bfseries Corresponding authors:}
Qinrui Liu$^{1*}$, Ali Kheirandish$^{2}$\\
{$^{1}$ \itshape Wisconsin IceCube Particle Astrophysics Center (WIPAC) and Department of Physics, University of Wisconsin-Madison, Madison, WI 53706, USA}\\
{$^{2}$ \itshape Department of Physics; Department of Astronomy \& Astrophysics; Center for Multimessenger Astrophysics, Institute for Gravitation \& the Cosmos, The Pennsylvania State University, University Park, PA 16802, USA}\\[4mm]
$^*$ Presenter

\FullConference{37$^{\rm{th}}$ International Cosmic Ray Conference (ICRC 2021)\\
		July 12th -- 23rd, 2021\\
		Online -- Berlin, Germany}
}

\begin{document}
\maketitle

\section{Introduction}
Cosmic rays (CRs) with energies up to several PeV, the "knee" in the CR spectrum, are believed to be of Galactic origin. However, where and how these CRs are accelerated remains an open question. Interactions of very high energy CRs in the Galaxy will lead to the production of pions, which subsequently decay into gamma rays and neutrinos, with energies reaching hundreds of TeV. As electromagnetic processes could also contribute to high-energy gamma-ray emission, only the detection of high-energy neutrinos would be a smoking gun for such CR interactions (i.e., hadronic interactions) as they are the only way to produce neutrinos. The sources of the vast majority of high-energy neutrinos detected by IceCube are yet to be identified. The isotropic distribution of high-energy neutrino's arrival direction suggests dominant contributions from extragalactic sources. The Galactic contribution to the diffuse neutrino flux is constrained to $\sim$14\% above 1 TeV \cite{Aartsen:2017ujz}. Studies have been conducted to identify Galactic point-like sources, extended regions, and the diffuse emission produced by CRs interacting with the interstellar medium. Nevertheless, recent searches for correlations do not show remarkable signals yet \cite{Albert:2018vxw,Kheirandish:2019bke,Aartsen:2020eof}. X-ray binaries (XRBs) are binary systems consisting of a compact object (neutron star (NS) or black hole (BH)) and a non-compact companion star. These systems are bright in X-rays and sometimes in gamma rays. XRBs have been proposed as sites of CR acceleration and hadronic interactions since the 1980s. XRBs with jets, often regarded as a smaller version of quasars and referred to as microquasars, have been widely discussed in the context of hadronic processes in jets. Protons can be accelerated in the jet, and pions are generated through interactions with the external radiation field of the accretion disk and/or internal synchrotron photons. Other discussions focus on hadronuclear interactions, e.g., jet-cloud/wind interactions when the jet is traversing the matter field of the ejected clouds or stellar wind from the companion star. For other XRBs where there is no collimated beam present, hadronic interactions can happen in a wider shocked region. CR acceleration can take place in the magnetosphere of a spinning NS and CRs can then further interact with matter from either the accretion disk or the companion star. See e.g.~\cite{Levinson:2001as,Anchordoqui:2002xu,Romero:2003td,Bednarek:2005gf} for theories of neutrino production in XRBs. Some XRBs have been observed at TeV energies, which illustrates the capability of these sources to accelerate particles to high energies.

 XRBs are known for their outburst and periodic emission. Thus, it is reasonable to hypothesize that the possible neutrino emission is related to either the periodicity or the X-ray outburst activity, which might stem from a change in the power or target material. Time-dependent analyses can be performed based on such hypotheses, which benefit from the suppression of the background, which is dominated by the atmospheric neutrino flux. Both time-integrated and time-dependent analyses searching for high-energy neutrino emission have been performed by IceCube and ANTARES, e.g., \cite{Aartsen:2015wto, Albert:2016gtl}, without significant detection. Here, we present a study focusing on XRBs using the IceCube muon track data searching for correlation with the X-ray outburst and persistent emission of the possible neutrino flux from XRBs, covering an ample list of sources.

\section{Analysis}
\begin{figure*}[hbt!]
    \centering
   \includegraphics[width=0.5\columnwidth]{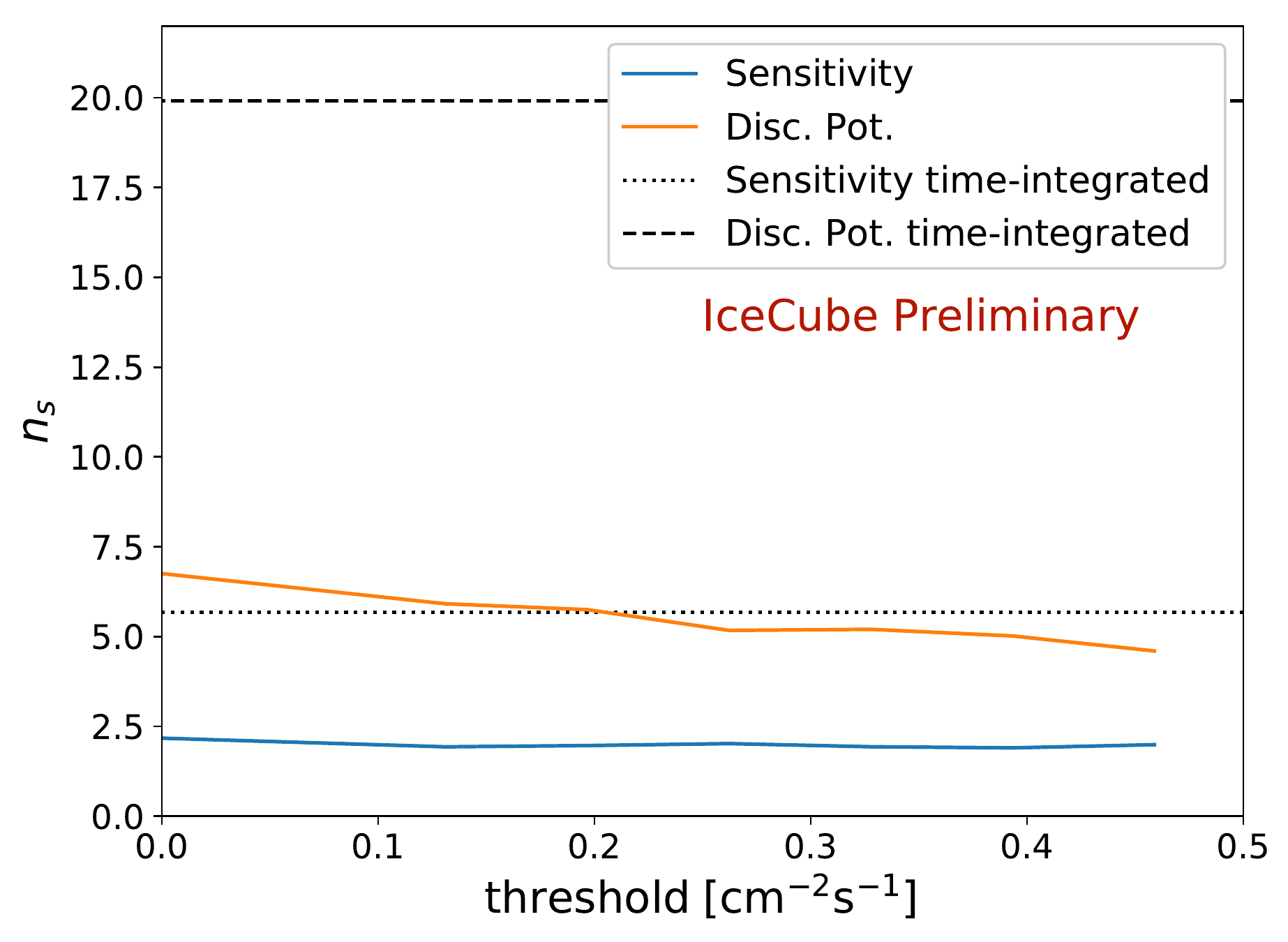}
    \caption{90\% sensitivity and 5$\sigma$ discovery potential of a flaring source V404 Cyg when varying the threshold (Bayesian blocked light curve in Fig.~\ref{fig:event_dist} ) and a comparison to the time-integrated case, which indicates an improvement in sensitivity. The spectrum shown here is $E^{-2}$.
    }
    \label{fig:sensitivity_V404}
\end{figure*}

This search uses an unbinned maximum likelihood method, which follows the one described in \cite{Braun:2008bg, Braun:2009wp}, to seek an excess of neutrino events (signal) above the background. In both the time-dependent and the time-integrated analyses, the likelihood function describing the signal includes both spatial clustering and energy information.  In the time-dependent analysis, a unique temporal term is incorporated in the likelihood, which incorporates a correlation between neutrinos and X-ray light curves. As the majority of the data is expected to be background events that are uniform in time, the likelihood function of the background is constructed with time-randomized data for the time-dependent analysis and right ascension randomized data for the time-integrated analysis. The test statistic is obtained by maximizing the likelihood function w.r.t. a set of parameters, which include the number of signal events ($n_s$) and the spectral index ($\gamma$) for both analyses. For the time-dependent analysis, in addition to $n_s$ and $\gamma$, time-related parameters introduced are the threshold of a light curve $f_{th}$ for picking flares and the time lag $T_{lag}$ between the X-ray and the neutrino emission.

The time-dependent analysis focuses on searching for a correlation between the neutrino emission and the X-ray activity of a source. For this purpose, hard X-ray light curves are used to construct the time probability density function (PDF). Light curves are obtained from hard X-ray data reported by \textit{Swift}/BAT in the energy range 15-50 keV \footnote{https://swift.gsfc.nasa.gov/results/transients/index.html}\citep{Krimm:2013lwa} and MAXI in the energy range 10-20 keV \footnote{http://maxi.riken.jp/top/slist.html} \citep{matsuoka2009maxi}. The X-ray light curve data are binned in days, and a Bayesian block algorithm is applied to find the optimal segmentation of the data and identify flares \cite{Scargle:2012gq}. After the light curves are divided into blocks, the value of each block can be fitted as a constant, taking into account the uncertainty of each data point. The normalized blocked light curves then act as the temporal PDF. Fig.~\ref{fig:sensitivity_V404} shows the sensitivity of the time-dependent analysis compared to the sensitivity of the time-integrated analysis from the direction of V404 Cyg, where the expected improvement is demonstrated.

The sources studied are from the Galactic high-mass XRB (HMXB) catalog \cite{Liu:2007tu} and the Galactic low-mass XRB (LMXB) catalog \cite{Liu:2007ts}, which include 301 sources. TeV sources from TeVCat \footnote{http://tevcat.uchicago.edu} \cite{wakely2008tevcat} which are not in the HMXB or LMXB catalog are added. Starting from the initial source list, sources without available \textit{Swift}/BAT or MAXI hard X-ray light curves are removed. As we are only interested in sources active with flaring or variable behaviors in X-ray, the variability and excess variance of the light curves are evaluated such that sources with weak emission are taken out. This step is applied only to the X-ray data in the time frame overlapped by the neutrino data sample. If both the \textit{Swift}/BAT and MAXI light curves pass the selection criteria, the \textit{Swift}/BAT data is preferred to be used for one source. After this selection, there are 102 sources in the initial source list left to be analyzed. 

\begin{figure*}[t!]
    \centering
   \includegraphics[width=1.0\columnwidth]{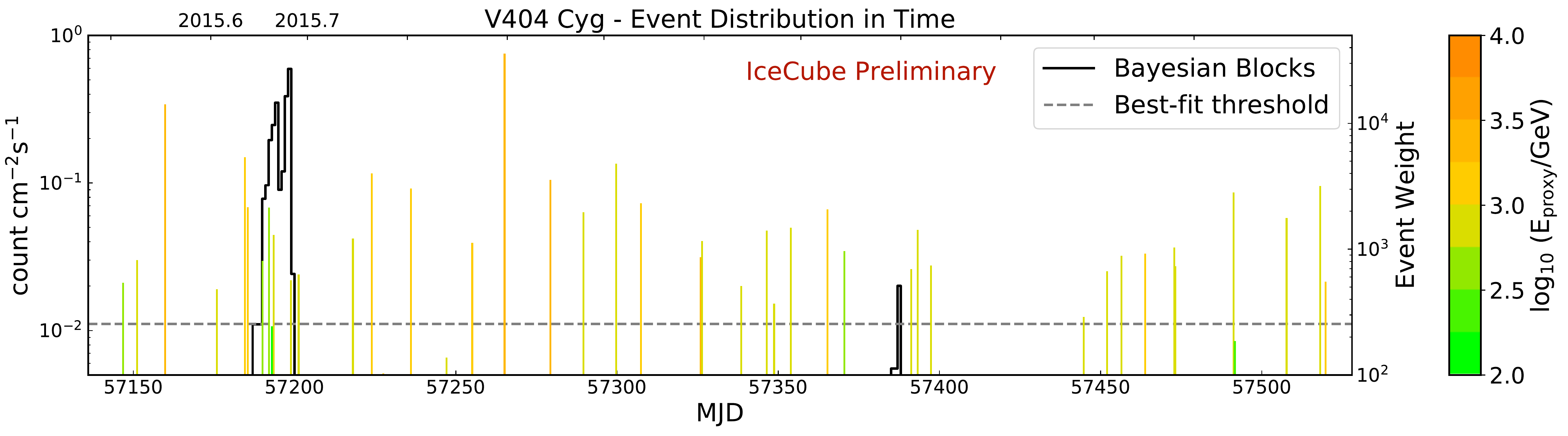}
    \caption{The temporal PDF before normalization (Bayesian blocks) and the event distribution within $1.5^\circ$ around V404 Cyg in the data sample of 2015 at the time indicated by MJD. The Bayesian blocks have been shifted by the best-fit time lag (-0.5 days) and the dashed gray line indicates the best-fit threshold (0.011~cm$^{-3}$\,s$^{-1}$). Vertical lines represent neutrino events. The color shows the energy proxy while the height shows the weight of each event in the likelihood function.
    }
    \label{fig:event_dist}
\end{figure*}
We complement the study with a time-integrated search for neutrino signals from four notable sources: Cyg X-3, LS 5039, LSI +61 303, and SS 433. Additionally, two time-integrated stacking tests are conducted for microquasars and TeV sources separately, with the method used in \cite{Aartsen:2020eof} and an equal weighting scheme when considering the relative contribution of each source. 

For all searches, we use 7.5 years of all-sky muon track data collected between 2011-05-13 and 2018-10-14, corresponding to a livetime of 2711 days. The data sample being used consists of high-quality through-going muon track events from the entire sky, yielding a total of 1502612 events. Details of the data sample are described in \cite{Aartsen:2016lmt}.

\section{Results \& Discussion}
\begin{table}[t!]
    \centering
    \begin{tabular}{ccccccc}
    \hline
    Analysis  & Name & TS & $\hat{n}_s$  & $\hat{\gamma}$   & $p$-value  & 90\% CL upper limits      \\
    \hline
     Flare  & V404 Cyg & 8.3 & 5.4  & 4.0 & 0.754$\;$(0.014)        & 0.91 \\
Time-integrated & Cyg X-3  & 6.8 & 44.6  & 3.3   & 0.036$\;$(0.009) &    1.51      \\
TeV XRB stacking & - &  0.1 & 7.7 & 3.5 & 0.587 & 1.22 \\
microquasar stacking & - & 0 & 0 & - & 1 & 7.32 \\
 \hline
    \end{tabular}
    \caption{The most significant source in the flare/time-integrated analysis with the TS, and the best-fitted $\hat{n}_s$ and  $\hat{\gamma}$. Both post-trial and pre-trial (bracketed) p-values are shown. The results of the 2 stacking tests are also listed. The 90\% CL upper limits are parameterized as $dN_{\nu_\mu+\bar{\nu}_\mu}/dE_\nu=\mathcal{F}_{\nu_\mu+\bar{\nu_\mu}}\left ( E_\nu/\rm{TeV} \right )^{-2}\,\cdot 10^{-4}\;\rm{TeV^{-1}cm^{-1}}$ for the flare analysis and $dN_{\nu_\mu+\bar{\nu}_\mu}/dE_\nu=\phi_{\nu_\mu+\bar{\nu_\mu}}\left ( E_\nu/\rm{TeV} \right )^{-2}\,\cdot 10^{-12}\;\rm{TeV^{-1}cm^{-1}s^{-1}}$ for time-integrated analyses. }
    \label{tab:results}
\end{table}
\hspace{-3cm}

\begin{figure}[t!]
    \centering
    \includegraphics[width=0.7\columnwidth]{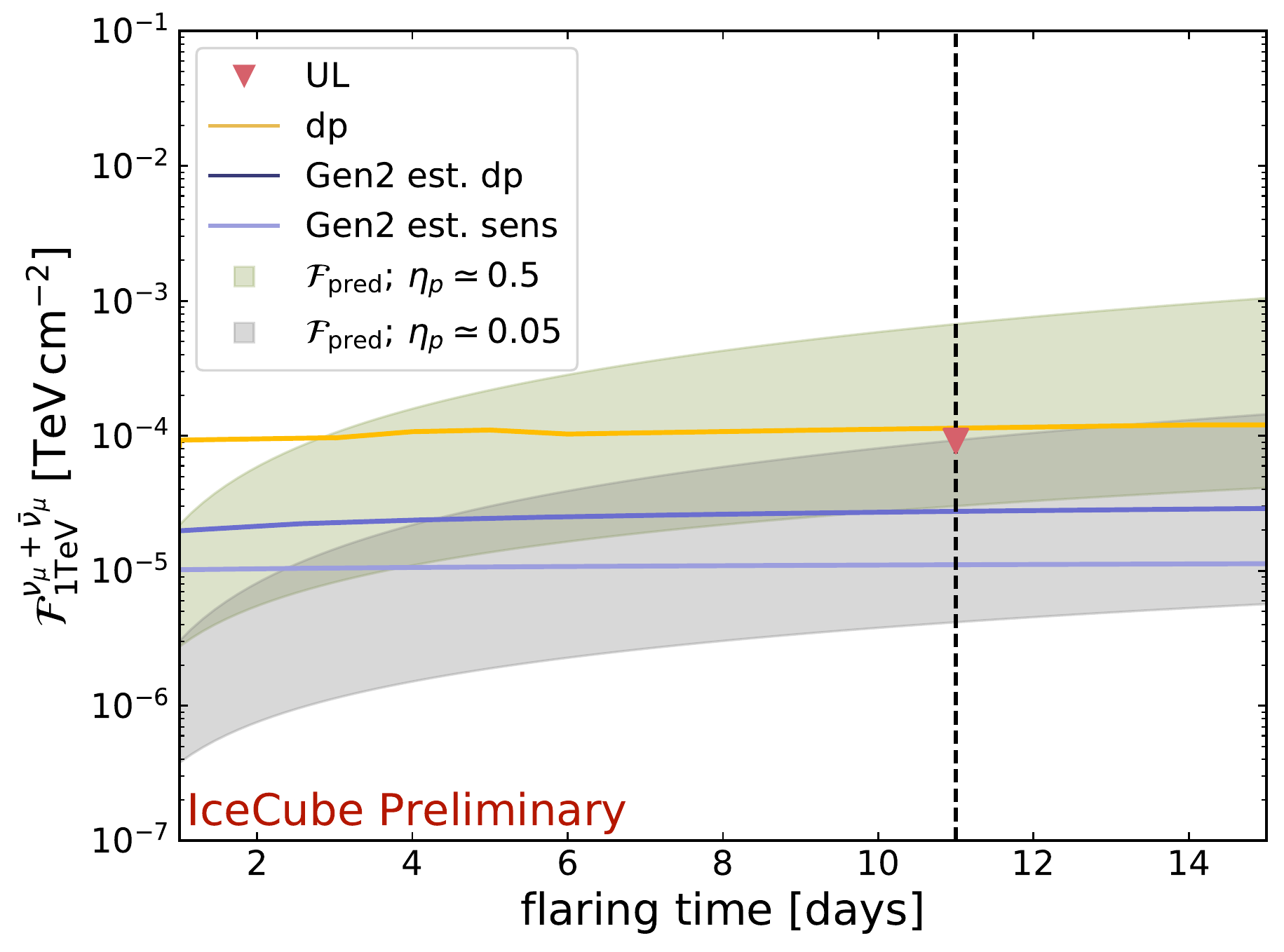} 
    \caption{The relation between the time-integrated flux at 1TeV and the flaring time for V404 Cyg. The dashed black line is the flaring time converted from the best-fit threshold and the red triangle shows the 90\% CL upper limit. The orange line is the 5$\sigma$ discovery potential in IceCube. Purple lines illustrate the estimated sensitivity at 90\% CL and 5$\sigma$  discovery potential in IceCube-Gen2. The shaded regions are the time-integrated neutrino flux prediction assuming an E$^{-2}$ spectrum with an energy cutoff at 100 TeV estimated following the jet model \citep{Distefano:2002qw}. The uncertainties are from flux densities in different frequencies in VLA radio measurements during the flaring time in 2015. The two colors correspond to varying the energy fraction of the jet carried by accelerated protons $\eta_p$.}
    \label{fig:V404}
\end{figure}

\begin{figure}[t!]
    \centering
    \includegraphics[width=0.7\columnwidth]{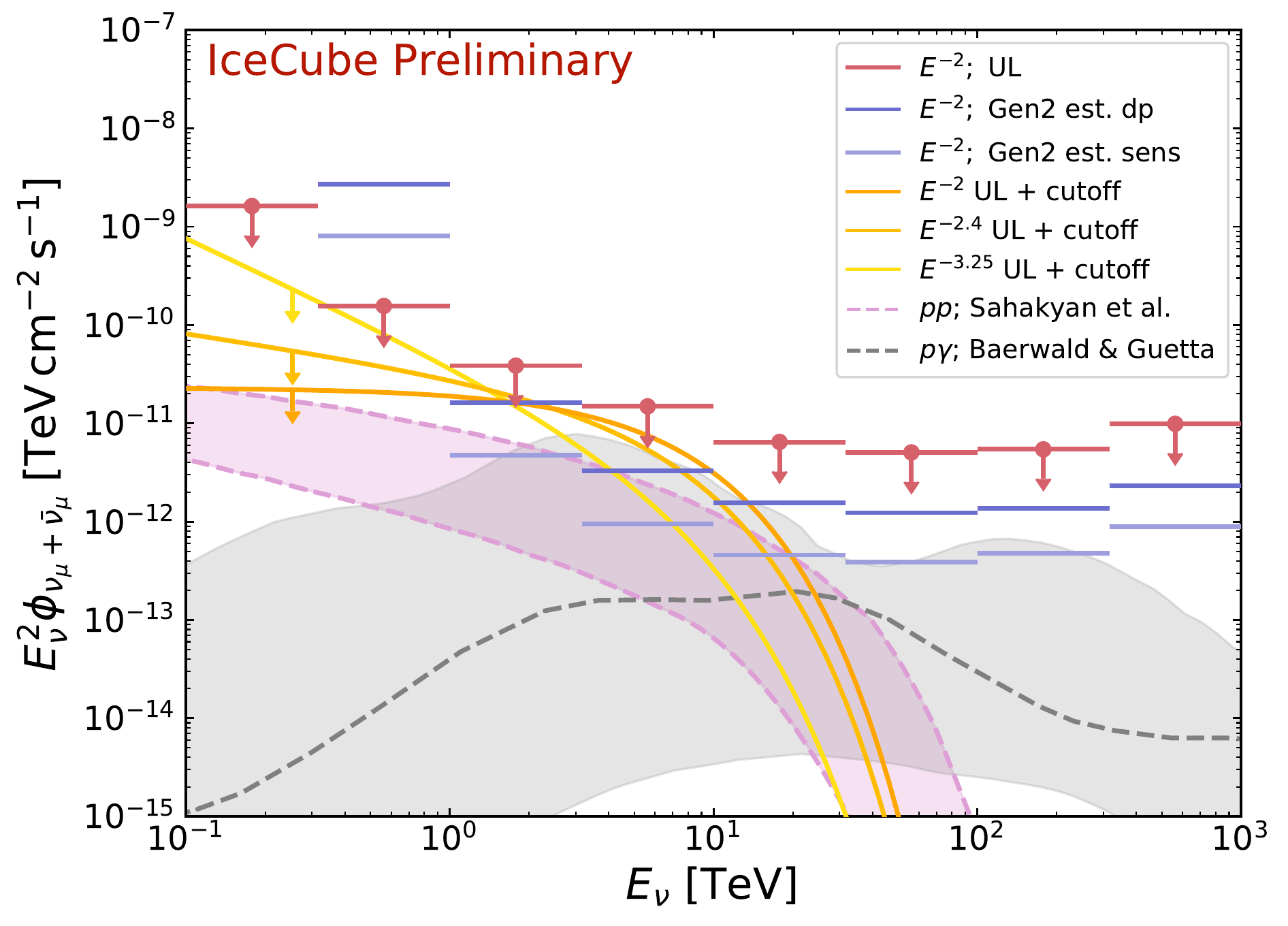} 
    \caption{ Red and purple lines indicate a comparison between current upper limits and estimated 10 yr sensitivity (light) \& discovery potential (dark) in IceCube-Gen2 for Cyg X-3. As the high-energy neutrino events nearby cut at several TeV, an exponential cutoff at 5 TeV is also applied for computing upper limits. The shaded regions show predictions of $pp$~\cite{Sahakyan:2013opa} and $p\gamma$~\cite{Baerwald:2012yd} scenarios. The inclusion of a cutoff is also to be compared to the shaded pink region which includes a cutoff of CR energy at 100 TeV with the spectral index ranging from 2.4-2.7. 3.25 corresponds to the best-fit spectral index. The gray shaded region shows the uncertainty from the collision radius.}
    \label{fig:CygX-3}
\end{figure} 

In the search for correlation of high-energy neutrinos and the flaring activity of XRBs,
the lowest p-value is found for the signal events from the microquasar V404 Cyg, a low-mass BH XRB, with a pre-trial p-value of 0.014. However, the $p$-value increases to 0.754 after taking into account the trials for the number of sources in the catalog. V404 Cyg underwent a major X-ray flaring episode in 2015. There are 5 sub-TeV neutrino events within $1.5^{\circ}$ of the source during the time of this flare, and the best-fit threshold indicates a time duration of 11 days, as shown in Fig.~\ref{fig:event_dist}. This giant flare was observed with a duration of approximately 13 days by \textit{Swift}/BAT \cite{segreto2015giant}.

In the time-integrated analysis, both the tests on individual sources and stacked search find no signal with sufficient statistical significance. The prominent excess in the point source search is found for Cyg X-3, which exhibits pre-trial p-value of $9\times 10^{-3}$, leading to a post-trial $p$-value 0.036 after considering the 4 trials. In the flare analysis, Cyg X-3 has a pre-trial $p$-value 0.09, less significant than the time-integrated results. Within $1^\circ$ around the source location, there are 44 events above 1~TeV, and the most energetic one among them has deposited energy about 5 TeV, leading to a soft best-fit spectrum. Since there is no significant signal found, we set the 90\% confidential level (CL) upper limits to the neutrino flux from the sources studied. A summary is shown in Table.~\ref{tab:results}. 

For microquasars, relativistic jets are expected to be the CR acceleration sites. Possible neutrino emission is expected from the beam dump on either radiation from the compact object itself or gas from the companion star. Parameters for neutrino flux prediction in \cite{Distefano:2002qw}, based on the photohadronic model of \cite{Levinson:2001as} can be constrained for some microquasars. Nevertheless, the simplified estimation has large uncertainties. For V404 Cyg, the X-ray flare in June 2015 was observed in multiple wavelengths, and the jet activity during that outburst was studied, e.g., in \cite{Miller-Jones:2019zla,Tetarenko:2018yrv}. A simple estimation of the neutrino flux using the jet model can be performed with the radio jet information when the source is in an outburst state. The upper limits reported here are compared to the time-integrated flux estimation in Fig.~\ref{fig:V404}. The collision region is estimated from the flaring duration. The values of jet parameters in the estimation are from \cite{Miller-Jones:2019zla,Tetarenko:2018yrv}, and the spectrum is assumed to be a power-law with an index of 2 and an exponential cutoff of 100 TeV.

For Cyg X-3, one of the microquasars identified as a gamma-ray source in early observations, many predictions have been calculated in the past decades depending on different models for microquasars. For a comparison to the upper limits, we take \cite{Baerwald:2012yd} and \cite{Sahakyan:2013opa}, which discussed the general $p\gamma$ and $pp$ scenarios based on the AGILE observation respectively, shown in Fig.~\ref{fig:CygX-3}. What needs to be mentioned is that Cyg X-3 lies in the direction of the Cygnus X region and is close to the Cygnus OB2 association but with a further distance compared to the Cygnus X region. The possibility of contamination from the Cygnus X complex cannot be excluded.

The next generation of the IceCube experiment, IceCube-Gen2, will provide a factor of eight increase in volume \cite{Aartsen:2020fgd}, leading to an expected $\sim$5-time increase in the effective area compared to IceCube, corresponding to an improvement in sensitivity by the same order, which advances the identification of neutrino sources. Here, we extend the study to IceCube-Gen2 and estimate the sensitivity and discovery potential for V404 Cyg, as an example of a flaring source and Cyg X-3, for persistently emitting sources. The estimated improvement can be seen in Fig.~\ref{fig:V404} and Fig.~\ref{fig:CygX-3}. The effective areas of muon tracks are computed from the proposed IceCube-Gen2 configuration, and the projection is evaluated similar to that in \cite{Aartsen:2020fgd} without considering a contribution from the existing IceCube detector. In comparison with theoretical calculations, it demonstrates the power to either identify those sources or rule out models with IceCube in the future.

\section{Summary}
A Galactic contribution to the high-energy neutrino flux observed by IceCube is expected. We present a study of neutrino emission from XRBs, long-standing candidates for the Galactic sources of CRs and neutrinos. We performed a time-dependent analysis based on the assumption of flaring neutrino emission. In parallel, a time-integrated search is also performed on 4 notable sources and 2 stacked lists. In the absence of any significant excess, we set upper limits on the neutrino emission in the scenarios discussed. The results of the most significant sources in this search are compared to models of neutrino production in XRBs. Our estimation of the improved detectability by IceCube-Gen2 due to higher neutrino event statistics demonstrates the potential for the future detection and presents a promising outlook of identifying Galactic cosmic-ray accelerators in the upcoming years.

\bibliography{references}
\bibliographystyle{ICRC}

%
%
%

\clearpage
\section*{Full Author List: IceCube Collaboration}




\scriptsize
\noindent
R. Abbasi$^{17}$,
M. Ackermann$^{59}$,
J. Adams$^{18}$,
J. A. Aguilar$^{12}$,
M. Ahlers$^{22}$,
M. Ahrens$^{50}$,
C. Alispach$^{28}$,
A. A. Alves Jr.$^{31}$,
N. M. Amin$^{42}$,
R. An$^{14}$
% ...
\\

\noindent
% ...
$^{12}$ Universit{\'e} Libre de Bruxelles, Science Faculty CP230, B-1050 Brussels, Belgium \\
% ...
$^{14}$ Department of Physics and Laboratory for Particle Physics and Cosmology, Harvard University, Cambridge, MA 02138, USA \\
% ...
$^{17}$ Department of Physics, Loyola University Chicago, Chicago, IL 60660, USA \\
$^{18}$ Dept. of Physics and Astronomy, University of Canterbury, Private Bag 4800, Christchurch, New Zealand \\
% ...
$^{22}$ Niels Bohr Institute, University of Copenhagen, DK-2100 Copenhagen, Denmark \\
% ...
$^{28}$ D{\'e}partement de physique nucl{\'e}aire et corpusculaire, Universit{\'e} de Gen{\`e}ve, CH-1211 Gen{\`e}ve, Switzerland \\
% ...
$^{31}$ Karlsruhe Institute of Technology, Institute for Astroparticle Physics, D-76021 Karlsruhe, Germany  \\
% ...
$^{42}$ Bartol Research Institute and Dept. of Physics and Astronomy, University of Delaware, Newark, DE 19716, USA \\
% ...
$^{50}$ Oskar Klein Centre and Dept. of Physics, Stockholm University, SE-10691 Stockholm, Sweden \\
% ...
$^{59}$ DESY, D-15738 Zeuthen, Germany \\
% ...

\subsection*{Acknowledgements}

\noindent
USA {\textendash} U.S. National Science Foundation-Office of Polar Programs,
U.S. National Science Foundation-Physics Division,
U.S. National Science Foundation-EPSCoR,
Wisconsin Alumni Research Foundation,
Center for High Throughput Computing (CHTC) at the University of Wisconsin{\textendash}Madison,
Open Science Grid (OSG),
Extreme Science and Engineering Discovery Environment (XSEDE),
Frontera computing project at the Texas Advanced Computing Center,
U.S. Department of Energy-National Energy Research Scientific Computing Center,
Particle astrophysics research computing center at the University of Maryland,
Institute for Cyber-Enabled Research at Michigan State University,
and Astroparticle physics computational facility at Marquette University;
Belgium {\textendash} Funds for Scientific Research (FRS-FNRS and FWO),
FWO Odysseus and Big Science programmes,
and Belgian Federal Science Policy Office (Belspo);
Germany {\textendash} Bundesministerium f{\"u}r Bildung und Forschung (BMBF),
Deutsche Forschungsgemeinschaft (DFG),
Helmholtz Alliance for Astroparticle Physics (HAP),
Initiative and Networking Fund of the Helmholtz Association,
Deutsches Elektronen Synchrotron (DESY),
and High Performance Computing cluster of the RWTH Aachen;
Sweden {\textendash} Swedish Research Council,
Swedish Polar Research Secretariat,
Swedish National Infrastructure for Computing (SNIC),
and Knut and Alice Wallenberg Foundation;
Australia {\textendash} Australian Research Council;
Canada {\textendash} Natural Sciences and Engineering Research Council of Canada,
Calcul Qu{\'e}bec, Compute Ontario, Canada Foundation for Innovation, WestGrid, and Compute Canada;
Denmark {\textendash} Villum Fonden and Carlsberg Foundation;
New Zealand {\textendash} Marsden Fund;
Japan {\textendash} Japan Society for Promotion of Science (JSPS)
and Institute for Global Prominent Research (IGPR) of Chiba University;
Korea {\textendash} National Research Foundation of Korea (NRF);
Switzerland {\textendash} Swiss National Science Foundation (SNSF);
United Kingdom {\textendash} Department of Physics, University of Oxford.
\end{document}